\documentclass[aps,tightenlines,showpacs,nofootinbib,twocolumn,floatfix]{revtex4}
\usepackage{epsfig}

\begin{document}

\title{Charmonium spectroscopy above thresholds}
\author{T. Fern\'andez-Caram\'es}
\affiliation{Departamento de F\'\i sica Fundamental, Universidad de Salamanca, E-37008
Salamanca, Spain}
\author{A. Valcarce}
\affiliation{Departamento de F\'\i sica Fundamental, Universidad de Salamanca, E-37008
Salamanca, Spain}
\author{J. Vijande}
\affiliation{Departamento de F\'{\i}sica At\'{o}mica, Molecular y Nuclear, Universidad de Valencia (UV)
and IFIC (UV-CSIC), Valencia, Spain.}
\date{\today}

\begin{abstract}
We present a systematic and selfconsistent analysis of four-quark
charmonium states and applied it to study compact four-quark systems
and meson-meson molecules. Our results are robust and should serve to
clarify the situation of charmonium spectroscopy
above the threshold production of charmed mesons.
\end{abstract}

\maketitle

Understanding of charmonium spectroscopy is challenging 
for experimentalists and theorists alike. 
Charmonium has been used as the test bed to demonstrate the color
Fermi-Breit structure of quark atoms obeying the same principles
as ordinary atoms~\cite{Isg83}.
Its nonrelativistic character 
($v/c \approx 0.2-0.3)$ gave rise to an amazing agreement between
experiment and simple quark potential model predictions
as $c\bar c$ states~\cite{Eic80}.
The opening of charmed meson thresholds was expected to modify
the trend in the construction of quark-antiquark models. 
In the adiabatic approximation
meson loops were absorbed into the static interquark potential. 
Thus, close to the threshold production of charmed mesons 
models required of an improved interaction ~\cite{Isg99}. 
The corrections introduced to the quark-antiquark spectra explained some deviations
observed experimentally \cite{Eic04}.

Since 2003, with the discovery of several states in the open charm sector,
we have witnessed a growth of puzzling new mesons, being $D_{sJ}^*(2317)$,
$D_{sJ}(2460)$ and $D_0^*(2308)$ the most prominent examples.
Later, several new states have joined this exclusive group either
in the open-charm sector, $D_{sJ}(2860)$ and $D_{sJ}(3040)$, or in the charmonium
spectra, like the well established $X(3872)$ and $Y(4260)$, $Z(3930)$,
$X(3940)$, $Y(3940)$, $X(4008)$, $X(4160)$, $X(4260)$, $Y(4350)$, and
$Y(4660)$. 
In addition, the Belle Collaboration has reported the observation
of similar states with non-zero electric charge: the $Z(4430)$, the
$Z_1(4040)$ and the $Z_2(4240)$ that have not yet been
confirmed by other experiments and remain somewhat controversial~\cite{Ols09}.
These new states do not fit, in general, the simple predictions of the quark-antiquark
schemes and, moreover, they overpopulate the expected number of states in (simple) two-body theories.
This situation is not uncommon in particle physics. For example,
in the light scalar-isoscalar meson sector hadronic molecules seem to
be needed to explain the experimental data~\cite{Jaf77,Wei90,Hoo08}. Also,
the study of the $NN$ system above the pion production threshold 
required new degrees of freedom to be incorporated in the theory, either
as pions or as excited states of the nucleon, i.e., 
the $\Delta$~\cite{Els88,Pop87}. 
This discussion suggests that charmonium spectroscopy could be rather
simple below the threshold production of charmed mesons but  
much more complex above it. In particular, the coupling to 
the closest $(c\bar c)(n\bar n)$ system, referred to 
as {\it unquenching the naive quark model}~\cite{Clo05},
could be an important spectroscopic ingredient. Besides,
hidden-charm four-quark states could 
explain the overpopulation of quark-antiquark theoretical states.
Thus, the new experimental discoveries are offering exciting 
new insights into the subtleties of the strong interaction.

In an attempt to disentangle the role played by multiquark configurations 
in the charmonium spectroscopy we have obtained an exact solution of the 
four-body problem based on an infinite expansion of the four-quark 
wave function in terms of hyperspherical harmonics~\cite{Vij07}. 
The method is exact but is not completely adequate to study states that are close to,
but below, the charmed meson production threshold. 
Such states are called {\it molecular}, in the sense
that they can be exactly expanded in terms of a single singlet-singlet
color vector. 
Close to a threshold, methods based on a series expansion fail to converge
since arbitrary large number of terms are required to determine the wave function.
From our analysis, we concluded that those four-quark states with two 
different asymptotic physical thresholds (as it is
the case of the $c\bar c n \bar n$ system that may split either
into a $(c\bar c)(n \bar n)$ or
$(c \bar n)(n \bar c)$ two-meson states) can hardly present a bound state
since the interaction between any pair of quarks contributes to the energy
of one of the two physical thresholds. 
However, we observed that the root mean square radius of a few channels
did not grow in the same manner as in those channels clearly converging
to an unbound two-meson threshold. Instead, their radius remained stable
and their energy did not cease slightly decreasing.

For this reason, we have used a different technique that we developed when studying
baryon spectra with screened potentials and that showed to be very powerful close to
a threshold \cite{Vij04}.
In this case, the hyperspherical
harmonic expansion of the wave function was computationally very expensive.
Instead, we solved the Faddeev equations for negative energies using the  
Fredholm determinant method that permitted us to obtain robust predictions even 
for zero-energy bound states.
For the charmonium the situation is similar but simpler. Similar because we
are working on the region where methods based on infinite expansions 
are inefficient, but simpler since it is a 
two-body problem, the scattering of two mesons. Thus, 
we solve the Lippmann-Schwinger equation 
looking for attractive channels that may contain a meson-meson molecule.
In order to account for all basis states we allow 
for the coupling to charmonium-light two-meson systems.
With this method we circumvent the uncertainties associated to the slow 
convergence of the hyperspherical harmonic method for large grand angular momenta.

When we consider the system of two mesons $M_1$ and $\overline{M}_2$ ($M_i=D, D^*$) 
in a relative $S-$state interacting through a potential $V$ that contains a
tensor force then, in general, there is a coupling to the 
$M_1\overline{M}_2$ $D-$wave and the
Lippmann-Schwinger equation of the system is 
\begin{eqnarray}
t_{ji}^{\ell s\ell^{\prime \prime }s^{\prime \prime }}(p,p^{\prime \prime };E)
&&\!\!\!\!\!=V_{ji}^{\ell s\ell^{\prime \prime }s^{\prime \prime }}(p,p^{\prime \prime
})+\sum_{\ell^{\prime }s^{\prime }}\int_{0}^{\infty }{p^{\prime }}%
^{2}dp^{\prime }\nonumber \\
\times V_{ji}^{\ell s \ell^{\prime }s^{\prime }}
(p,p^{\prime }) &&\!\!\!\!\!\!\!
{\frac{1}{E-{p^{\prime }}^{2}/2{\bf \mu }+i\epsilon }}%
t_{ji}^{\ell^{\prime }s^{\prime }\ell^{\prime \prime }s^{\prime \prime
}}(p^{\prime },p^{\prime \prime };E),  \label{eq1}
\end{eqnarray}
where $t$ is the two-body amplitude, $j$, $i$, and $E$ are the
angular momentum, isospin and energy of the system, and $\ell s$,
$\ell^{\prime }s^{\prime }$, $\ell^{\prime \prime }s^{\prime \prime }$
are the initial, intermediate, and final orbital angular momentum
and spin; $p$
and $\mu $ are the relative momentum and reduced mass of the
two-body system, respectively. 
In the case of a two $D$ meson system that can couple to a charmonium-light
two-meson state, for example when $D\overline{D}^*$ is coupled to
 $J/\Psi \omega$, the Lippmann-Schwinger equation for
$D\overline{D}^*$ scattering becomes
\begin{eqnarray}
\!\!\!\!\!\!\!\!t_{\alpha\beta;ji}^{\ell_\alpha s_\alpha \ell_\beta s_\beta}(p_\alpha,p_\beta;E)&&\!\!\!\!\!\! = 
V_{\alpha\beta;ji}^{\ell_\alpha s_\alpha \ell_\beta s_\beta}(p_\alpha,p_\beta)+ \nonumber \\
\!\!\sum_{\gamma=D\overline{D}^*,J/\Psi\omega}\sum_{\ell_\gamma=0,2} 
&&\!\!\!\!\!\!  \int_0^\infty p_\gamma^2 dp_\gamma
V_{\alpha\gamma;ji}^{\ell_\alpha s_\alpha \ell_\gamma s_\gamma}
(p_\alpha,p_\gamma) \nonumber \\
\!\! \times \, G_\gamma(E;p_\gamma)
&&\!\!\!\!\!\!  t_{\gamma\beta;ji}^{\ell_\gamma s_\gamma \ell_\beta s_\beta}
(p_\gamma,p_\beta;E),
\label{eq2}
\end{eqnarray}
with $\alpha, \beta= D\overline{D}^*, J/\Psi \omega$.
For bound-state $E < 0$ that the singularity of the propagator
is never reached, we can neglect $i\epsilon$ in the denominator.
By changing variables,
\begin{equation}
p'(p_\gamma) = b{1+x' \over 1-x'},
\label{eq3}
\end{equation}
where $b$ is a scale parameter, and the same for $p (p_\alpha)$ 
and $p" (p_\beta)$. Replacing the integral from $-1$ to 1 by a
Gauss-Legendre quadrature, we obtain a set of linear equations.
If a bound state exists at an energy $E_B$,
the matrix determinant is zero.
We took the scale parameter $b$ of Eqs.~(\ref{eq1}) and~(\ref{eq2}) to be
$b$ = 3 fm$^{-1}$
and used a Gauss-Legendre quadrature with $N$ = 20 points.  

We have consistently used the same interacting Hamiltonian to study
the two- and four-quark systems to guarantee that thresholds
and possible bound states are eigenstates of the same Hamiltonian. 
Such interaction contains a universal
one-gluon exchange, confinement, and a chiral potential between
light quarks~\cite{Vij05}. We have solved the coupled channel problem of the 
$D\overline D$, $D\overline{D}^*$, and $D^*\overline{D}^*$. In all cases
we have included the coupling to the relevant $(c\bar c)(n\bar n)$
channel (from now on denoted as $J/\Psi \omega$ channels).

As we study systems with well-defined 
$C-$parity, 
let us comment on the $D\overline{D}^*$ system.
Since neither $D\overline{D}^*$ nor $\overline{D}D^*$ 
are eigenstates of $C-$parity, it
is necessary to construct the proper linear combinations. Taking into account that
$C(D) = \overline{D}$ and $C(D^*) = - \overline{D}^*$ (with a relative minus sin
between them), it can be found that~\cite{Bra07}:
\begin{equation}
D_1 = \frac{1}{\sqrt{2}} \left( D\overline{D}^* + \overline{D}D^*\right)
\end{equation}
and
\begin{equation}
D_2 = \frac{1}{\sqrt{2}} \left( D\overline{D}^* - \overline{D}D^*\right)
\end{equation}
are the eigenstates corresponding to $C = -1$ and $C = +1$, respectively.
This does not depend on the quantum numbers of the system because $D$ and
$D^*$ are not a particle-antiparticle pair. Once the $C-$parity of a 
$D\overline{D}^*$ state is fixed, its isospin is also determined. 
In particular, for a $D\overline{D}^*$ $S-$wave state, positive $C-$parity requires
isospin 0, while negative $C-$parity implies isospin 1.

\begin{figure}[t]
\caption{Fredholm determinant for the $J^{PC}(I)=1^{++}(0)$ $D\overline{D}^*$
system. Solid (dashed) line: results with (without) coupling to the $J/\Psi \omega$
channel.}
\mbox{\epsfxsize=85mm\epsffile{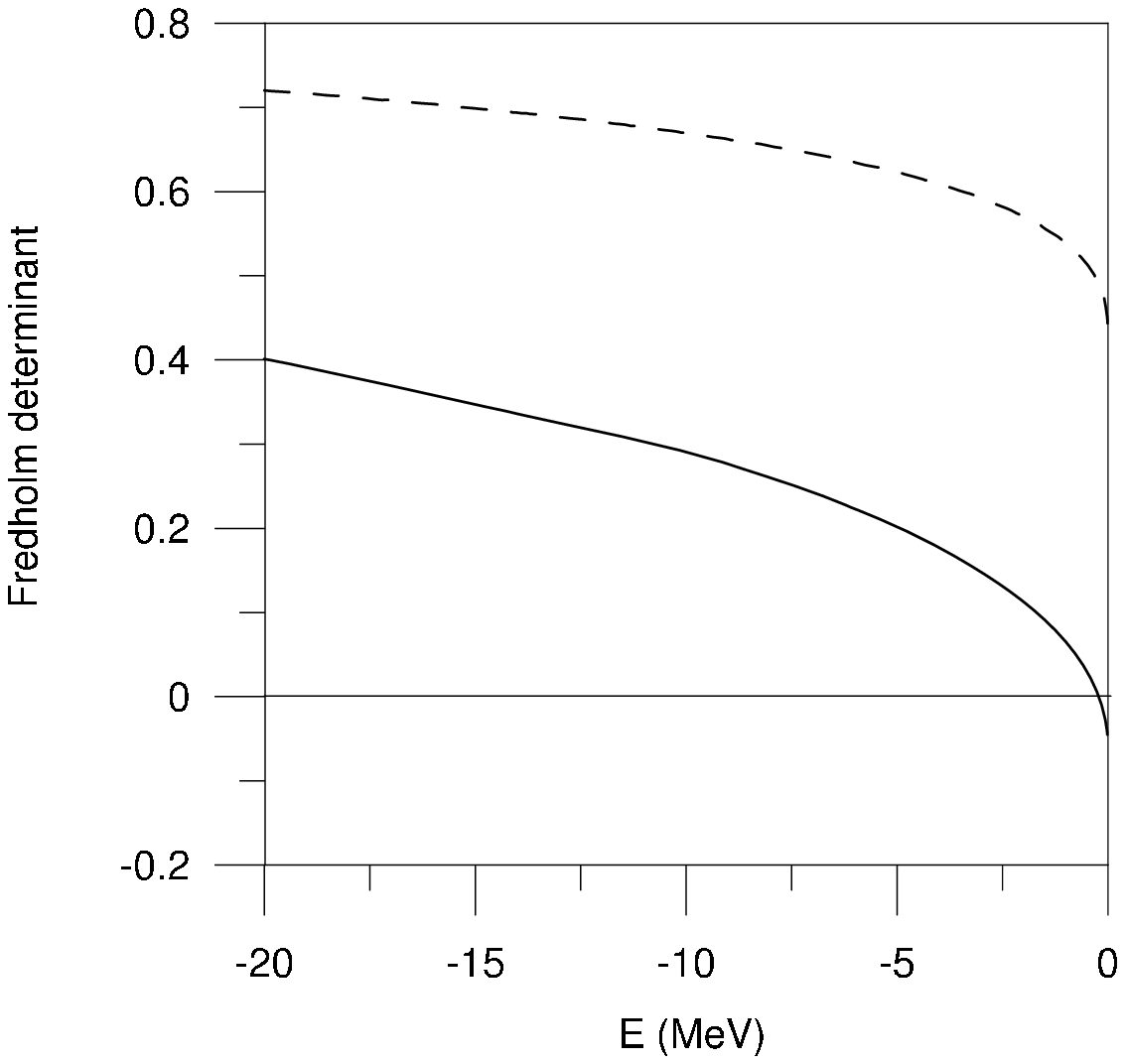}}
\label{f1}
\end{figure}

\begin{figure}[b]
\caption{Fredholm determinant of the most attractive $J^{PC}(I)$
channels for the $D\overline D$ and $D^*\overline{D}^*$ systems.}
\mbox{\epsfxsize=85mm\epsffile{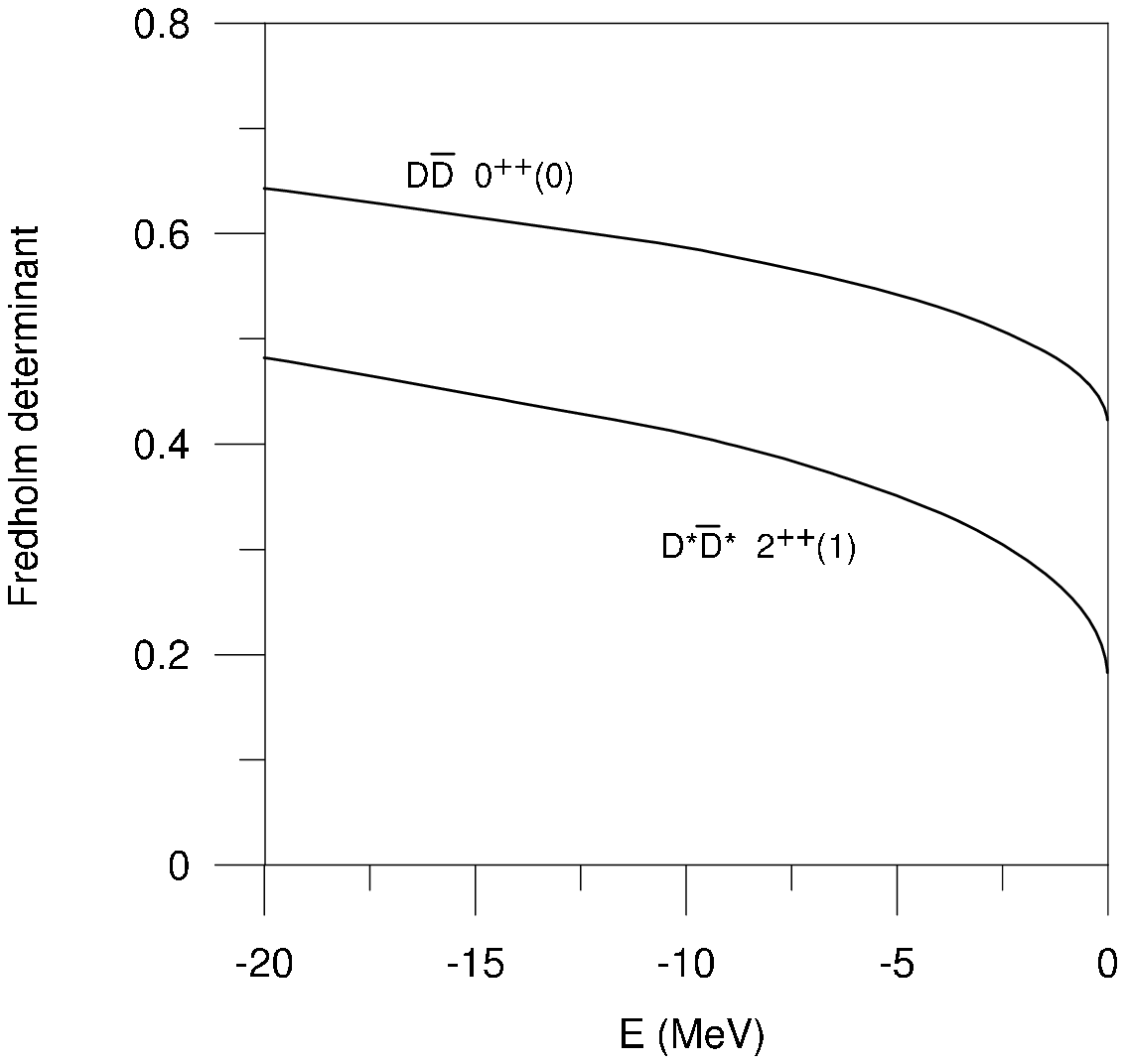}}
\label{f2}
\end{figure}

Table~\ref{t1} and Figs.~\ref{f1} and~\ref{f2} summarize our results.
We have specified the quantum numbers and plotted the Fredholm
determinant of the attractive channels. 
The rest, not shown on the table, are either repulsive or have zero probability to 
contain a bound state or a resonance.
Let us remark that, of all possible channels, only a few are attractive.
Of the 
systems made of a particle and its corresponding antiparticle,
the $J^{PC}(I)=0^{++}(0)$ channel is always attractive. In general,
the coupling to the $\eta_c \eta$ channel
reduces the attraction, but there is still enough attraction to expect a resonance 
close and above the threshold. This channel is much more attractive
for the $D^*\overline{D}^*$ system than for $D\overline{D}$, thus, in the latter 
one could expect a wider resonance. 
It is easy to explain the reason for such a close-to-bind situation 
with these quantum numbers. They
can be reached from a two-meson system without explicit orbital
angular momentum, while through a simple $c\bar c$ pair
it needs a unit of orbital angular momentum. 
Similar arguments were used to explain the proliferation
of light scalar-isoscalar mesons ~\cite{Jaf77,Wei90,Hoo08}.
The most attractive channel in the $D\overline{D}^*$ case
is the $J^{PC}(I)=1^{++}(0)$ and can be explained as before, except 
the unity of intrinsic spin due to the $D^*$ meson.
A simple calculation of the $D\overline{D}^*$ system (Eq.~(\ref{eq1})) indicates that
the $J^{PC}(I)=1^{++}(0)$ and $1^{+-}(1)$ are degenerate. It is
the coupling to the $J/\Psi \omega$ (Eq.~(\ref{eq2})) that breaks the degeneracy to
make the $1^{++}(0)$ more attractive. 
The isospin 1
channel becomes repulsive due to the coupling to the lightest channel that 
includes a pion. Then, the existence of
meson-meson molecules in the isospin one $D\overline{D}^*$ channels can be discarded.
Using the coupling to the $J/\Psi \omega$, not present
in the calculations at the hadronic level of~\cite{Liu08,Tho08},
we obtain a
binding energy for the $J^{PC}(I)=1^{++}(0)$ in the range $0-1$ MeV,
in good agreement with the experimental measurements of $X(3872)$ (see Fig.~\ref{f1}).
This result supports
the analysis of the Belle data on $B \to K + J/\Psi \pi^+ \pi^-$ and
$B \to K + D^0 \overline{D}^0 \pi^0$ that favors the $X(3872)$ being
a bound state whose mass is below the $D^0\overline{D}^0$ 
threshold~\cite{Bra07}. The existence of a bound state
in the $1^{++}(0)$ $D\overline{D}^*$ channel would not show up
in the $D\overline{D}$ system because of quantum
number conservation.

Finally, we have found that the $J^{PC}(I)=2^{++}(0,1)$ $D^*\overline{D}^*$
(see Fig.~\ref{f2}) are also attractive due to the coupling to the $J/\Psi \omega$ and 
$J/\Psi \rho$ channels, respectively. This would give rise to new
states around 4 GeV/c$^2$ and one 
experimental candidate could be the $Y(4008)$.
In this case, such a resonance would also appear
in the $D\overline{D}$ system for large relative
orbital angular momentum, $L=2$. A
similar behavior can be observed in resonances predicted
for the $\Delta \Delta$ system~\cite{Val01}.
\begin{table}[tbp]
\caption{Attractive channels for the two $D-$mesons system.}
\label{t1}
\begin{tabular}{|c|c|}
\hline
System & $J^{PC}(I)$ \\
\hline\hline
$D\overline{D}$ & $0^{++}(0)$ \\ 
\hline
$D\overline{D}^*$ & $1^{++}(0)$ \\ 
\hline
$D^*\overline{D}^*$ & $0^{++}(0)$ \\ 
$D^*\overline{D}^*$ & $2^{++}(0)$ \\ 
$D^*\overline{D}^*$ & $2^{++}(1)$ \\ 
\hline
\end{tabular}
\end{table}
 
In all cases, being loosely bound states whose masses are
close to the sum of their constituent meson masses,
their decay and production properties must be quite 
different from conventional $q\bar q$ mesons.
Our calculation does not exclude a possible mixture of
standard charmonium states in the channels where
we have found attractive molecular systems. 
This admixture could explain some properties of the
$X(3872)$~\cite{Ger06,Big09}. 
We would like to emphasize the similarity of our results
to those of Ref.~\cite{Tor91} in spite 
of our different approach. Our treatment is general, dealing
simultaneously with the two- and four-body problems and using
an interaction containing gluon and quark exchanges instead of
the simple two-body one-pion exchange potential of Ref.~\cite{Tor91}.
Nevertheless, we also concluded that
the lighter meson-meson molecules are in
the vector-vector and pseudoscalar-vector two-meson channels. 
Finally, let us remark that our approach could also be applied to the
the $c\bar cs\bar s$ sector.

To summarize, we have performed the first systematic analysis of four-quark
hidden-charm states as compact states or meson-meson molecules. For the first
time we have performed a consistent study of all quantum numbers within the
same model. 
Our predictions robustly show that no deeply bound states can be expected for this system.
Only a few channels can be expected to 
present observable resonances or slightly bound states. Among them, we
have found that the $D\overline{D}^*$ system must show a bound state
slightly below the threshold for charmed mesons production
with quantum numbers $J^{PC}(I)=1^{++}(0)$,
that could correspond to the widely discussed $X(3872)$. Of the systems 
made of a particle and its corresponding antiparticle,
$D\overline{D}$ and $D^*\overline{D}^*$, the $J^{PC}(I)=0^{++}(0)$
is attractive. It would be the only candidate to accommodate
a wide resonance for the $D\overline{D}$ system. 
For the $D^*\overline{D}^*$ the attraction
is stronger and structures may be observed close and above the charmed
meson production threshold.
Also, we have shown that the $J^{PC}(I)=2^{++}(0,1)$ $D^*\overline{D}^*$
channels are attractive due to the coupling to the $J/\Psi \omega$ and 
$J/\Psi \rho$ channels. 
Due to heavy quark symmetry, replacing the charm
quarks by bottom quarks decreases the kinetic energy without significantly
changing the potential energy. In consequence, four-quark bottomonium mesons
must also exist and have larger binding energies. An experimental effort
in this direction will confirm or rule out the theoretical expectations.
If the scenario presented here turns out to be correct, it will open a new
interesting spectroscopic area.

When this work was finished we learned that particular studies of
some of the new charmonium states coincide with our theoretical predictions
about the more attractive quantum numbers~\cite{Gut09}.

\acknowledgments
The authors thank D.~Jenkins, V.~Vento, F.~Atrio-Barandela, L.~Garc\'\i a
and J.C.~Pav\'on
for suggestions and/or a careful reading of the manuscript.
This work has been partially funded by the Spanish Ministerio de
Educaci\'on y Ciencia and EU FEDER under Contract No. FPA2007-65748,
by Junta de Castilla y Le\'{o}n under Contracts No. SA016A17 and GR12, 
and by the Spanish Consolider-Ingenio 2010 Program CPAN (CSD2007-00042),

\end{document}